\documentclass[twocolumn,prl,superscriptaddress,showpacs]{revtex4}
\usepackage{color}
\usepackage{amsmath}
\usepackage[english]{babel}
\usepackage{amssymb}
\usepackage{graphicx}
\usepackage{esint}

\begin{document}
\title{Vacuum-induced Autler-Townes splitting in a superconducting artificial atom}
\author{Z.H. Peng}\email{zhihui_peng@riken.jp}
\affiliation{Key Laboratory of Low-Dimensional Quantum Structures
and Quantum Control of Ministry of Education, Department of Physics
and Synergetic Innovation Center for Quantum Effects and
Applications, Hunan Normal University, Changsha 410081, China}
\affiliation{Center for Emergent Matter Science, RIKEN, Wako, Saitama 351-0198, Japan}

\author{J.H. Ding}
\affiliation{Institute of Microelectronics, Tsinghua University, Beijing 100084, China}
\author{Y. Zhou}\affiliation{Department of Physics, Tokyo University of Science, Kagurazaka, Tokyo 162-8601, Japan}
\author{L.L. Ying}\affiliation{Center for Excellence in
Superconducting Electronics,
Shanghai Institute of Microsystem and Ibformation Technology, Chinese Academy of Sciences, Shanghai 200050, China}
\author{Z. Wang}\affiliation{Center for Excellence in
Superconducting Electronics,
Shanghai Institute of Microsystem and Ibformation Technology, Chinese Academy of Sciences, Shanghai 200050, China}
\author{L. Zhou}\affiliation{Key Laboratory of Low-Dimensional Quantum Structures
and Quantum Control of Ministry of Education, Department of Physics
and Synergetic Innovation Center for Quantum Effects and
Applications, Hunan Normal University, Changsha 410081, China}
\author{L.M. Kuang}\affiliation{Key Laboratory of Low-Dimensional Quantum Structures
and Quantum Control of Ministry of Education, Department of Physics
and Synergetic Innovation Center for Quantum Effects and
Applications, Hunan Normal University, Changsha 410081, China}
\author{Yu-xi Liu}\email{yuxiliu@mail.tsinghua.edu.cn}
\affiliation{Institute of Microelectronics, Tsinghua University, Beijing 100084, China}
\affiliation{Tsinghua National Laboratory for Information Science and Technology (TNList), Beijing 100084, China}
\author{O. Astafiev}\email{Oleg.Astafiev@rhul.ac.uk}
\affiliation{Physics Department, Royal Holloway, University of London, Egham, Surrey TW20 0EX, United Kingdom}\affiliation{National Physical Laboratory, Teddington, TW11 0LW, United Kingdom}
\affiliation{Moscow Institute of Physics and Technology, Dolgoprudny, 141700, Russia}
\author{J.S. Tsai}
\affiliation{Department of Physics, Tokyo University of Science, Kagurazaka, Tokyo 162-8601, Japan}
\affiliation{Center for Emergent Matter Science, RIKEN, Wako, Saitama 351-0198, Japan}

\begin{abstract}
We experimentally study a vacuum-induced Autler-Townes doublet in a superconducting three-level artificial atom strongly coupled to a coplanar waveguide resonator and simultaneously to a transmission line. The Autler-Townes splitting is observed in the reflection spectrum from the three-level atom in a transition between the ground state and the second excited state when the transition between the two excited states is resonant with a resonator. By applying a driving field to the resonator, we observe a change in the regime of the Autler-Townes splitting from quantum (vacuum-induced) to classical (with many resonator photons). Furthermore, we show that the reflection of propagating microwaves in a transmission line could be controlled by different frequency single photons in a resonator.
\end{abstract}
\maketitle

Electromagnetic waves propagating through a medium of identical atoms are resonantly absorbed.
The absorption can be eliminated when a strong driving field couples other atomic transitions in, for example, three-level atoms, creating a transparency window for the waves. This results in either Autler-Townes splitting (ATS)~\cite{Autler1955} or electromagnetically induced transparency (EIT)~\cite{Harris1990,Harris1997} and is extensively studied in quantum optics~\cite{Fleischhauer2005,Scully-book}. The same phenomena can be observed even if the medium is replaced by a single atom or a molecule which is coupled to either a cavity~\cite{Muller2007,Mucke2010} or an open space~\cite{Tey2008,Hwang2009}. However, the strong coupling between the driving field and a single natural atom is difficult to achieve. Recently, the strong coupling has been experimentally achieved between a superconducting artificial atom and non-quantized microwave fields confined in a one-dimensional transmission line~\cite{Astafiev2010}. This enables ATS and EIT to be observed using a single superconducting artificial atom~\cite{Murali2004,Ian2010,Sun2014,Gu2016}. Several experiments have already shown ATS~\cite{Baur2009,Sillanpaa2009,Abdumalikov2010,Li2012,Hoi2013PRL,Hoi2013,Novikov2013,Liu2017} and state manipulation~\cite{Kelly2010,Xu2016} in a single three-level artificial atom. ATS~\cite{Suri2013} and EIT~\cite{Sovikov2015} have been demonstrated in a Jaynes-Cummings ladder system of a single two-level artificial atom dispersively coupled to a cavity.

\begin{figure}
\center
\includegraphics[scale=0.4]{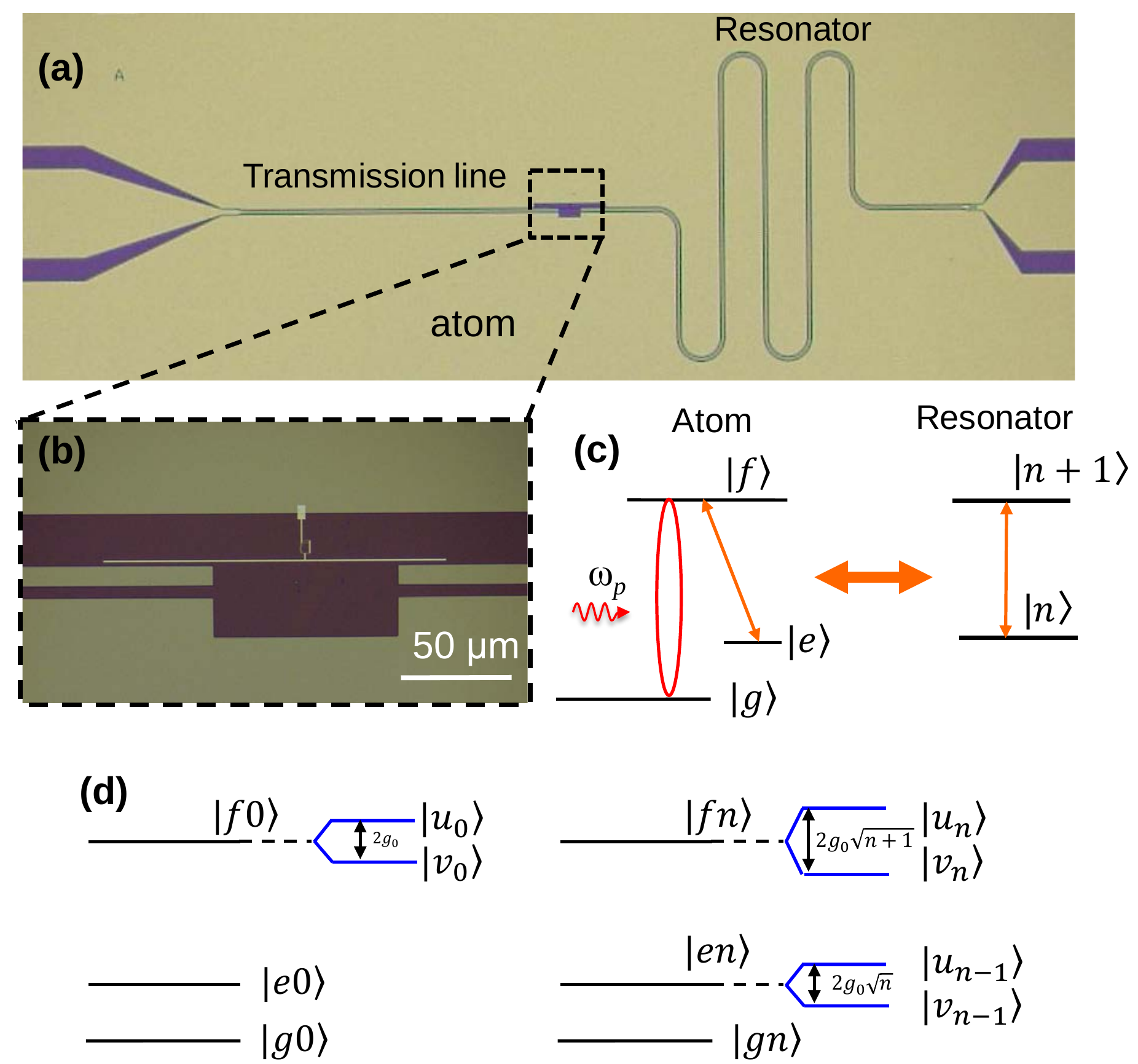}
\caption{({a}) Micrograph of the device, comprising a superconducting artificial atom capacitively coupled to a transmission line resonator (meandering structure). Additionally, the artificial atom is coupled to a transmission line (on the left side) used to directly measure the reflection spectrum from the atom.
({b}) Magnified micrograph of the artificial atom of the tunable-gap superconducting flux qubit geometry. ({c}) Schematic of the three-level artificial atom coupled to a resonator. The transition from the ground state to the second excited state is probed at frequency $\omega_p$.
({d}) Left panel: The dressed state picture of the vacuum induced (quantized) ATS due to coupling to a resonator in $|e\rangle \leftrightarrow |f\rangle$. Right panel: The dressed state picture of the classical ATS with many resonator photons. }
\label{picture}
\end{figure}

The observation of both ATS and EIT in a medium (an ensemble of atoms) usually requires strong classical driving fields. However, the conditions are different: the driving field required for the observation of EIT is weaker than that for ATS, and there is a crossover from ATS to EIT~\cite{Sanders,Yang}. In particular, in ATS the splitting is determined by the Rabi frequency of the driving field, whereas in EIT the dip between the two peaks is a result of quantum interference. These effects take place even if the medium is scaled down to a single atom and the driving field is quantized. The condition for the observation of the effects can then be reduced to a small number of photons or even without photons owing to coupling to a vacuum mode. Theoretical investigations show that the transparency for the classical probing field can still occur~\cite{Field1993}. This has potentially important applications, for example, in single-photon switches and transistors, all-optical quantum logic, quantum communication, and metrology~\cite{Chang2014}. However, vacuum-induced transparency has only recently been observed in an ensemble of three-level atoms~\cite{Tanji-Suzuki2011}. Here, we demonstrate ATS with a single three-level artificial atom, which is controlled by a quantized single-mode field in a transmission line resonator.

Our device, shown in Fig.~\ref{picture}(a), consists of a three-level artificial atom capacitively and strongly coupled to two macroscopic objects simultaneously: a coplanar waveguide resonator (CPWR) and a transmission line. Although the device is reminiscent of the one studied in Ref.~\cite{Peng2016}, where the atom was coupled to two transmission lines, it is essentially different. We emphasize that in spite of the strong coupling of the atom to the transmission line and the atom to the resonator, the line and resonator are decoupled from each other (details are given in Supplementary Material ~\cite{Suppl}). This is a peculiarity of the device owing to the unique property of superconducting quantum systems: they are micrometer-scale electronic circuits. Namely, the two macroscopic objects (i.e. open lines, resonators) can be effectively decoupled from each other but, at the same time, strongly coupled to the quantum circuit. The artificial atom has the geometry of a tunable-gap flux qubit: a superconducting loop with two Josephson junctions and a dc-SQUID ($\alpha$-loop)~\cite{Fedorov2010,Peng2016}, shown in Fig.~\ref{picture}(b), is fabricated near the voltage antinode (open end) of the CPWR using electron-beam lithography and Al/AlO$_x$/Al shadow evaporation (see the SEM image in~\cite{Suppl}).  A weak probe signal with frequency $\omega_p$ is applied from the transmission line (left-hand side). Both the half-wavelength CPWR and the transmission line are etched from a $50\,\rm nm$ thick Nb thin film sputtered on an oxidized undoped silicon wafer. The lines have a center conductor of $10\,\mu\rm m$ width separated from the ground planes by $6\,\mu\rm m$ gaps, resulting in a $50\,\Omega$ wave impedance.

Our experiment is carried out in a dilution refrigerator with a base temperature of about 30~mK. The fundamental frequency of the CPWR is $\omega_r=2\pi\times8.794\,$GHz with the decay rate $\kappa=2\pi\times3.6\,$MHz. The three lowest energy levels of the artificial atom, denoted by $|g\rangle$ for the ground state,  $|e\rangle$ for the first excited state, and $|f\rangle$ for the second excited state, are controlled by the magnetic flux $\Phi$ in the loop. The minimum transition energies from the ground to both excited states occur at half-integer flux quanta $\Phi_{N}=(N+\frac{1}{2})\Phi_{0}$, where $N$ is an integer and $\Phi_{0}$ is the flux quantum. The minimal energy gap $\Delta$ for the transition from $|g\rangle$ to $|f\rangle$ can be tuned by controlling the magnetic flux $\Phi_{\alpha}$ penetrating through the $\alpha$-loop. When the biased flux $\Phi$ is close to $\Phi_{N}$, the two lowest-energy eigenstates ($|g\rangle$ and $|e\rangle$) are in superposition of the two oppositely circulating persistent current states of the loop. The energy between the two lowest levels $\hbar\omega_{ge}$ in the vicinity of $\Phi_{N}$ is approximated by the expression $\sqrt{(2I_{p}\delta\Phi)^2+\Delta(\Phi_{N})^2}$, where $\delta\Phi=\Phi-\Phi_{N}$ and $I_{p}$ is the persistent current in the qubit loop. The weak dependence of $\Delta$ on $\Phi$ can be safely neglected when $\delta\Phi\ll\Phi_0$, which means that $\Delta(\Phi)\approx\Delta(\Phi_{N})$. As shown in Fig.~\ref{picture}(c), by choosing $\Phi_{N}$ and $\delta\Phi$, we can tune both transition frequencies $\omega_{ef}$ and $\omega_{gf}$. The ATS can be observed by measuring directly the reflection spectrum through the left transmission line.

As shown in Fig.~\ref{picture}(c), the transition energy between the $|e\rangle$ and $|f\rangle$ states is aligned to the resonator resonance. The probe field is applied at the $|g\rangle$ to $|f\rangle$ transition. Thus, the effective Hamiltonian of the whole system driven by classical waves with amplitude $\Omega$ at frequency $\omega_{p}$ close to the $|g\rangle \leftrightarrow |f\rangle$ transition is given by
\begin{eqnarray} \label{effectHami}
H &=&\hbar \omega_{r} a^{\dag}a+\hbar\omega_{gf} \sigma_{ff}+\hbar \omega_{ge}\sigma _{ee} \\  \nonumber
&&+\hbar g_{0}\left( a^{\dag}\sigma _{ef}+a\sigma _{fe}\right) + 2\hbar \Omega(\sigma_{fg}+\sigma_{gf}) \cos\omega_p t.
\end{eqnarray}
Here, the atomic operator $\sigma_{jk}$ is defined as $\sigma_{jk}=|j\rangle\langle k|$ with $\{|j\rangle,|k\rangle\}=\{|g\rangle,|e\rangle, |f\rangle\}$ and $a^\dag$ and $a$ are the photon creation and annihilation operators in the single-mode resonator, respectively.
The first four terms of Eq. (1) represent the Jaynes-Cummings Hamiltonian for the two-level system with the following specific features: it presents the interaction of two excited states ($|e\rangle$ and $|f\rangle$) with the resonator rather than the ground and excited states. Nevertheless, the Jaynes-Cummings physics can be applied here~\cite{JC1963,Haroche2006,Rempe1987}.
Namely, in the vicinity of the $|e\rangle \leftrightarrow |f\rangle$ resonance with the resonator, $|f\rangle$ level splits due to the interaction with the resonator vacuum mode, which can be described in terms of zero-photon dressed states as $|v_{0}\rangle=\cos\theta|f0\rangle-\sin\theta|e1\rangle$ and $|u_{0}\rangle=\sin\theta|f0\rangle+\cos\theta|e1\rangle$, where $\tan{2\theta} = 2g_{0}/(\omega_{r}-\omega_{f}+\omega_{e})$ and the state $|jn\rangle=|j\rangle\otimes|n\rangle$ with the two quantum numbers $j$ and $n$, denoting the three-level atomic states $|j\rangle$ and the Fock states of resonator $|n\rangle=\lbrace|0\rangle,|1\rangle,|2\rangle,...\rbrace$, respectively. The dressed states are schematically shown in the left panel of Fig.~\ref{picture}(d), while the right panel shows the general case of a system with $n$ photons in the resonator. The last term represents the probing field which couples the ground and the second excited states ($|g\rangle \leftrightarrow |f\rangle$, where the vacuum mode $|0\rangle$ is omitted for simplification).

\begin{figure}
\center
\includegraphics[scale=0.45]{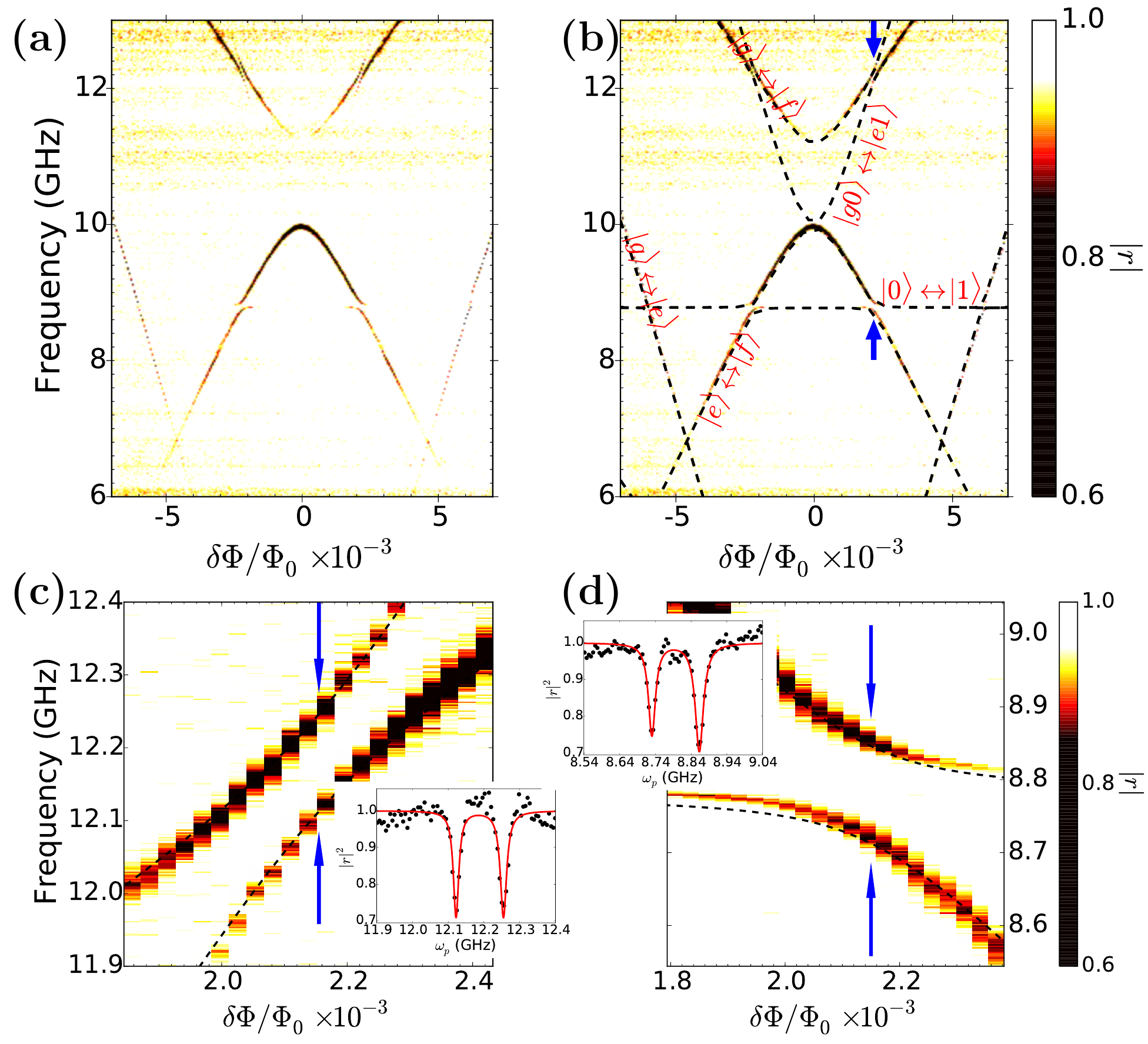}
\caption{(a) Spectrum of the artificial atom vs biased flux plotted as a reflection coefficient $|r|$. The transition $|e\rangle \leftrightarrow |f\rangle$ is visible due to the thermal population of the $|e\rangle$ level. (b) The spectrum with theoretically calculated energy levels (dashed lines) from the full system Hamiltonian. At $\delta\Phi\sim2.15\times10^{-3}\Phi_{0}$ (indicated by arrows), the quantum regime of ATS (vacuum splitting) is observed due to resonant coupling between the $|e\rangle \leftrightarrow |f\rangle$ transition and the resonator. (c) The anticrossing in the $|g\rangle \leftrightarrow |f\rangle$ spectroscopy line. (d) The anticrossing measured in the transition $|e\rangle \leftrightarrow |f\rangle$. The splitting sizes are essentially the same because both are determined by the atom-resonator coherent interaction strength $g_{0}$. The insets of (c) and (d) show the splitting at the positions indicated by arrows. }
\label{ATS}
\end{figure}

In our experiment, we first preset the flux bias to $\Phi_{N} = 0.5\Phi_{0}$ ($N$ = 0) and characterize the energy bands of our system by measuring the reflection with a vector network analyzer. A weak probe signal is applied to the transmission line from the left-hand side in Figs.~\ref{picture}(a-b) and scanned in the frequency range from $6$ to $12.5$ GHz. By measuring the reflection amplitude $r$ versus the probe frequency $\omega_p$ and a narrow range of flux bias $\delta\Phi$ ($\delta\Phi \ll \Phi_0$), the transitions $|g\rangle \leftrightarrow |e\rangle$ and $|g\rangle \leftrightarrow |f\rangle$ are revealed as sharp dips. This is presented in Fig.~\ref{ATS}(a) as reflection coefficient $|r|$ -- the normalized reflection amplitude.
The transition frequencies of the atom at $\delta\Phi = 0$ reach minima $\omega_{ge}=2\pi\times1.3\,\rm GHz$ and $\omega_{gf}=2\pi\times11.2\,\rm GHz$, respectively (the data below $2\pi\times6\,$GHz are not shown).
In addition to the transitions from the ground state, we are able to directly measure the transitions between excited states $|e\rangle$ and $|f\rangle$. This is possible because of the finite population of the first excited state $|e\rangle$ due to the non-equilibrium background radiation from the control lines. The $|e\rangle \leftrightarrow |f\rangle$ line is `brightest' at the degeneracy point ($\delta\Phi = 0$), where $\hbar\omega_{eg}$ is minimal. It is difficult to avoid the unwanted population of the $|e\rangle$ level with an energy gap of about 1~GHz, but in our experiment it is useful as it allows to reveal the full spectroscopy picture. The calibrated system effective temperature is around 75~mK (details in Supplementary Material ~\cite{Suppl}), which is similar to the results of other groups \cite{Geerlings2013,Jin2015}. The transitions $|g\rangle \leftrightarrow |e\rangle$ and $|e\rangle \leftrightarrow |f\rangle$ are bright due to large transition matrix elements. However, the $|g\rangle \leftrightarrow |f\rangle$ and $|g0\rangle \leftrightarrow |e1\rangle$ lines vanish at $\delta\Phi = 0$ as they are prohibited due to selection rules~\cite{Abdumalikov2010,Inomata2012}.
The dashed lines in Fig.~\ref{ATS}(b) represent the calculations from the full-system Hamiltonian with Josephson energy $E_{J}=h \times242\,\rm GHz$ of the junctions, charging energy $E_{c}= h \times4.2\,\rm GHz$, and the geometrical factor $\alpha=0.8$ of the SQUID Josephson junctions~\cite{Inomata2012}. Our artificial atom is coupled to the resonator through capacitance $C_{c} \approx 2.0\, \rm fF$. 

Note that the $|e\rangle \leftrightarrow |f\rangle$ transition between the two excited states is resonant with the resonator at $\delta\Phi\sim2.15\times10^{-3}\Phi_{0}$. 
Next, we measure the reflection from the system in the vicinity of the resonance $\omega_{ef} \approx \omega_r$.  The result ($|r|$ in the frequency range from 6 to 12.5~GHz vs $\delta\Phi$) is shown in Figs.~\ref{ATS}(c) and (d). The anticrossings are clearly observed in the two spectroscopic lines corresponding to the transitions $|g\rangle \leftrightarrow |f\rangle$ in Fig.~\ref{ATS}(c) and $|e\rangle \leftrightarrow |f\rangle$ in Fig.~\ref{ATS}(d), respectively. This confirms the strong coupling between the excited state transition ($|e\rangle \leftrightarrow |f\rangle$) of the atom and the resonator vacuum mode ($|0\rangle \leftrightarrow |1\rangle$). The line shapes at the exact resonance, indicated by blue arrows in Figs.~\ref{ATS}(c) and (d), are presented in the insets. The splitting sizes are $2g_{0}=2\pi\times132\,$MHz because the observed effect is the quantized ATS due to the coupling between transition $|e\rangle \leftrightarrow |f\rangle$ and the vacuum mode of the resonator. We estimate the probing amplitude to be $\Omega/2\pi \approx$ 7 MHz.

After confirming the vacuum-induced ATS, we study the dependence of the ATS on the resonant driving to the resonator field at $\omega_{r} \approx 2\pi\times$~8.794~GHz using a probe signal at $\omega_{p} \approx 2\pi\times$~12.2~GHz, which corresponds to the transition between the ground state and the second excited state ($|g\rangle \leftrightarrow |f\rangle$). The dynamics of this driven atom-resonator system can be described by the Markovian master equation \cite{Ding2017}
\begin{eqnarray} \label{MasterEQ}
\dot{\rho} &=&-\frac{i}{\hbar}[H_{F},\rho]+\kappa(2a\rho a^{+}-a^{+}a\rho-\rho a^{+}a)\\ \nonumber
& &+\sum_{j,k=g,e,f}^{j\geq k}\frac{\gamma_{jk}}{2}[2\sigma_{kj}\rho\sigma_{jk}-\sigma_{jk}\sigma_{kj}\rho-\rho\sigma_{jk}\sigma_{kj}],
\end{eqnarray}
for the density matrix $\rho$ of the coupled system, where $\gamma_{jk}$ is the relaxation/dephasing rate of the atom.
Here, $H_{F}=H+\hbar \Omega_{d}(a^{+}e^{-i\omega_{r}t}+ae^{i\omega_{r}t})$ is the Hamiltonian when the driving with amplitude $\Omega_d$ to the cavity field is considered and the Hamiltonian $H$ is given by Eq.(\ref{effectHami}).

\begin{figure}
\center
\includegraphics[scale=0.85]{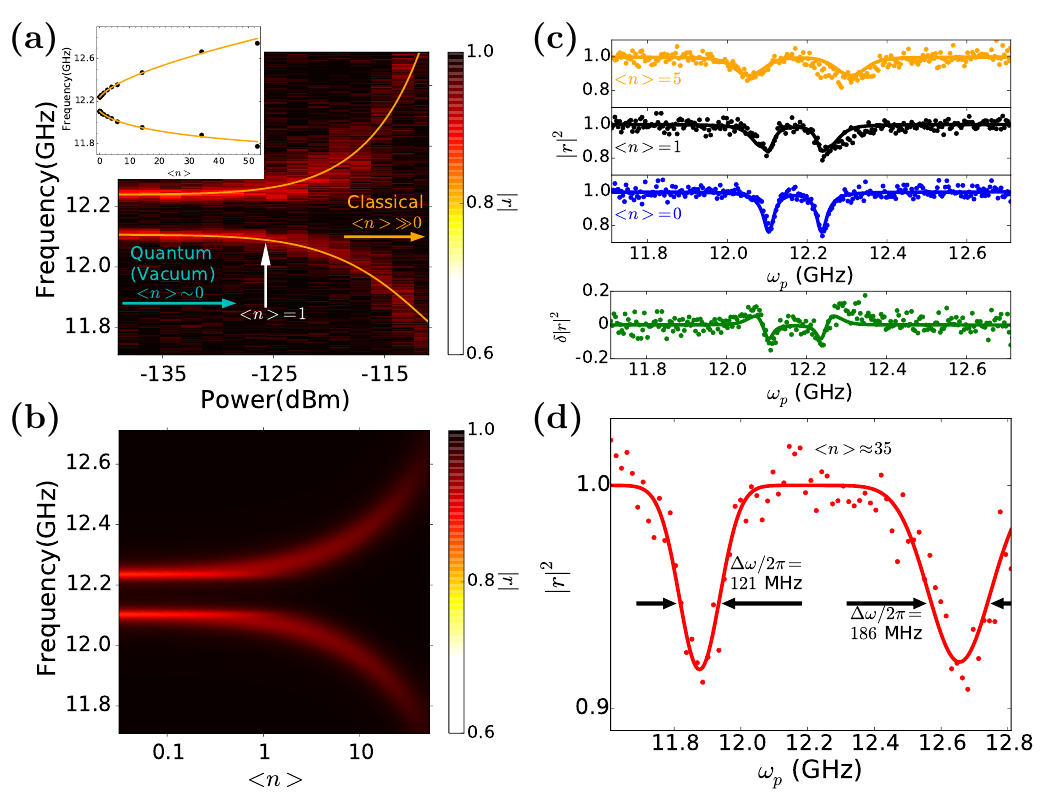}
\caption{(a) Dependence of ATS on driving power of the resonator. The orange guidelines show the splitting. The inset shows dip positions versus calibrated photon numbers. The solid lines are fits accounted the ac Zeeman shift. (b) Simulated ATS in the three-level atom coupled to a resonator. (c) Observed splittings at $\langle n\rangle = 0$ (quantum regime), $\langle n\rangle = 1$ and $\langle n\rangle = 5$ (upper three panels). The solid lines show the results of calculations. The bottom panel shows the difference between the traces with $\langle n\rangle= 1$ and $\langle n\rangle = 0$. (d) The splitting at $P = -113$~dBm, corresponding to $\langle n\rangle \approx 35$. }
\label{ATSVsPower}
\end{figure}

The reflection coefficient versus frequency is shown in Fig.~\ref{ATSVsPower}(a), where the $x$-axis is the driving power in the resonator with the probing amplitude $\Omega/2\pi \approx 7$~MHz for the $|g\rangle \leftrightarrow |f\rangle$ transition close to the weak driving limit \cite{Suppl}. In this regime, the splitting is determined by the coupling to the vacuum mode of the resonator. The vacuum-induced ATS is a quantum mechanical phenomenon, which cannot be observed in the classically driven ATS. At larger power, the splitting starts to increase according to $2 g_{0}\sqrt{\langle n\rangle + 1}$, where $\langle n\rangle$ is the mean photon number inside the resonator. We call quantum regime the one, where the vacuum splitting gives the main contribution to the ATS. Note that there are other quantum effects that can be observed beyond the subject of the present work (e.g., resolving photon states due to quantized fields in the resonator). The asymmetry of the branches is explained by the ac Zeeman shift (or ac Stark shift) \cite{Abdumalikov2008,Schuster2005} from the $|g\rangle \leftrightarrow |f\rangle$ transition dispersively coupled to the resonator, and the frequency shift $\zeta \langle n\rangle$ is found to be $\zeta$ = 2.5 MHz per photon. The mechanism is different from that of the asymmetric ATS observed in superconducting qubits with weak anharmonicity~\cite{Baur2009,Sillanpaa2009,Li2012}. The extracted dip positions versus calibrated photon number $\langle n\rangle$ inside the resonator are presented in the inset. The solid lines show the calculated dip positions $\omega_{n}$ from the relations $\omega_{n}=\omega_{fg}\pm g_0 \sqrt{\langle n\rangle+1} + \zeta \langle n\rangle$ including the correction of the ac Zeeman shift.

The simulated of the reflection coefficient is shown in Fig.~\ref{ATSVsPower}(b) with a simplified model by numerically solving Eq. (\ref{MasterEQ}) in a steady state with the artificial atom parameters: $\omega_{eg}=2\pi\times3.379\,$GHz,  $\omega_{fg}=2\pi\times12.173\,$GHz, $\gamma_{eg}=2\pi\times2\,$MHz, $\gamma_{fg}=2\pi\times4\,$MHz, $\gamma_{fe}=2\pi\times25\,$MHz, $\gamma_{ff}=2\pi\times1\,$MHz, and $\gamma_{ee}=2\pi\times20\,$MHz. The dephasing effect comes from $|f\rangle \leftrightarrow |g\rangle$ fluctuations, which induced by the photon shot noise inside the cavity, is also added to $\gamma_{ff}$~\cite{Schuster2005}. The other second-order effects in the $|f\rangle \leftrightarrow |g\rangle$ interaction with the resonator are neglected. We simplify our model for the numerical calculations because of limited computer resources insufficient to operate with the full Hamiltonian in the dissipative regime with many photons because of the too large Hilbert space. Note that although there is a weak signature of zero-one state photon splitting, it cannot be resolved in the experiment because of the measurement noise. In our experiment, it is not yet reach the reliable resolution of the two states, which can be expressed as $\Delta\omega_{10} > 2\lambda_{fg}$ -- the half-spacing between two dips is larger than $\lambda_{fg}$ which is the decay rate of the atomic off-diagonal terms $\rho_{fg}$, where $\Delta\omega_{10}=g_0 (\sqrt{2}-1) = 2\pi\times$~27~MHz and $\lambda_{fg}=2\pi\times$~16~MHz(details are given in~\cite{Suppl}).

Figure~\ref{ATSVsPower}(c) shows the dependence of reflection power with $\langle n\rangle = 0$, $\langle n\rangle = 1$ and $\langle n\rangle = 5$. The bottom panel shows the difference between the traces with $\langle n\rangle = 0$ and $\langle n\rangle = 1$. It demonstrates the reflection of propagating microwaves in the transmission line could be controlled by microwave fields in the resonator at the single-photon level. 
It may find potential applications in single-photon switches, single-photon transistors, and so on~\cite{Chen2013,Shomroni2014,Baur2014,Zhou2013}.

The intrinsic linewidth broadening with increasing driving power when $\langle n\rangle \gg 1$ is explained by the dephasing of $|f\rangle \leftrightarrow |g\rangle$ from the fluctuations of the photon inside the cavity~\cite{Schuster2005}. This is contrast to the linewidth narrowing with increasing driving power in ATS driven with classical control fields~\cite{Abdumalikov2010}.
The upper branch of the spectrum is broadened in the vicinity of 12.4~GHz, probably due to an unwanted low quality mode in the environment, which is difficult to remove at high (above 10~GHz) frequencies. Figure 3(d) shows the reflection power when $\langle n\rangle \approx 35$. The red line is fitted by a double-Gaussian curve of the form $A\exp{(-2\big(\delta\omega/\Delta\omega}\big)^2)$, where the fitting parameter $\Delta\omega/2\pi$ is found to be $121$ MHz at 11.876~GHz and $186$~MHz at 12.655~GHz. The linewidths are about twice the coupling strength $g_0$, which can be mainly determined by the Poisson photon distribution of coherent states formed inside the resonator~\cite{Schuster2005}.

Finally, we change the coupling strength $g_{0}$ between the transition $|e\rangle \leftrightarrow |f\rangle$ and the resonator by tuning the magnetic field inside the device $\alpha$-loop \cite{Fedorov2010}. We confirm that the splitting in the reflection spectrum of the coupled system around the transition frequency $\omega_{fe}$ is also twice the coupling strength $g_{0}$.

In conclusion, we engineer an interaction between a superconducting artificial atom and a resonator and observe the vacuum-induced (quantized) Autler-Townes doublet by direct measurement of the reflection spectrum from the atom. Furthermore, we study the transition of the quantum-to-classical Autler-Townes doublet with changing the number of photons inside the resonator.
Our work is an important step towards the control of propagating microwaves in a transmission line by a single photon confined in a different frequency resonator.
With the further improvements, the effect may have potential applications in single-photon switches, single-photon transistors, and so on.

\begin{acknowledgements}
This work was funded by ImPACT Progam of Council for Science, Technology and Innovation (Cabinet Office, Government of Japan) and NEDO IoT program. O.~V.~Astafiev is supported by Russian Science Foundation, grant No.~16-12-00070.  Z.~W. is supported by the Strategic Priority Research Program (B) of the Chinese Academy of Sciences (XDB04010100 and XDB040104000). L.~M.~Kuang is supported by the National Natural Science Foundation of China under Grants Nos.11775075 and 11434011, the National Fundamental Research Program of China (the 973 Program) under Grant No. 2013CB921804. L. Zhou is supported by the National Natural Science Foundation of China under Grants Nos.1142250 and 11374095. Y.~X. Liu is supported by NSFC under Grant No. 91321208 and National Basic Research Program of China Grant No. 2014CB921401.
\end{acknowledgements}

\end{document}